# Dielectric properties of strained NiO thin films

Alireza Kashir[1*], Hyeon-Woo Jeong[1], Gil-ho Lee[1], Pavlo Mikheenko[2], Yoon Hee Jeong[1*]

[1]Department of Physics, Pohang University of Science and Technology (POSTECH), Pohang, 37673, Republic of Korea

[2]Department of Physics, University of Oslo, P.O. Box 1048 Blindern, 0316 Oslo, Norway

## Abstract

The dielectric properties of NiO thin films grown by pulsed laser deposition have been studied as a function of strain at temperature from 10 to 300 K. Above 150 K, the contribution of space-charge polarization to the dielectric permittivity of NiO films becomes dominant, and the more defective films, which were grown at low temperatures shows a drastical increase in the dielectric constant up to room temperature. While the atomically-ordered film, which was grown at high temperature doesn't show any considerable change in the dielectric constant in the range from 10 to 300 K. Below 100 K, the effect of strain on the dielectric constant becomes clear. An increase in dielectric permittivity is observed in the strained films while the relaxed film doesn't show any remarkable deviation from its bulk value. The low-temperature dielectric behavior of NiO thin films can be interpreted based on the effect of strain on the lattice dynamics of rocksalt binary oxides.

**Keywords:** Dielectric properties; Space-charge polarization; NiO; Pulsed laser deposition; Strained thin films;

Corresponding Authors:

*yhj@postech.ac.kr
Tel: +82-54-279-2078

*kashir@postech.ac.kr

## 1. Introduction

The effect of thickness, strain and defect structure on dielectric properties and lattice dynamics of the strongly correlated systems has been playing crucial role in the development of solid-state electronics. During the last two decades, by the advancement of the precise methods of fabrication and characterization of the high-quality thin films, the investigation of electronic properties of the materials under diverse conditions became real and revealed several interesting phenomena. Some of them were predicted theoretically and emerged after application of appropriate strain, controlled introduction of defects and reduction in the film thickness [1–6].

Due to their relatively simple crystal and electronic structure, cubic rocksalt magnetic binary oxides have been a model system to investigate the emergent phenomena under the complex conditions. During the last decade, they have attracted a lot of attention, especially in theoretical condensed matter physics since using first principle calculations some non-trivial phenomena have been predicted [7–11].

Nickel oxide (NiO) (the only stable oxide in Ni-O binary system) crystalizes in cubic rocksalt structure with the lattice constant of 4.177 Å. It belongs to the wide band-gap (~ 3.6-4 eV) type-II antiferromagnets with the Neel temperature of 523 K. Below $T_N$ it undergoes a cubic to rhombohedral phase transition (α=90.06°) due to the exchange striction which is originated from magnetic exchange interaction at the onset of the magnetic alignment [11]. Stoichiometric NiO is a strong electrical insulator (ρ=$10^{13}$ ohm.cm), but its resistivity decreases drastically with introduction of the point defects (especially Ni deficiencies) or at a small-concentration Li doping [12]. For a long time, NiO has been the benchmark in the development of solid-state physics. It shows diverse interesting phenomena under different condition, like spin-phonon coupling at Neel temperature [13], insulator-metal transition under high pressure [14,15], anomalous magnetic behavior in nanoparticle structures [16], ferromagnetism induced by the point defects [17] etc.

Dielectric properties of NiO ceramics have been studied theoretically and experimentally [18–21] under different conditions and the value of dielectric constant $ε_r$ of 11.9 ($D = εE$, $ε = ε_r ε_0$, where $D$ is the electric displacement field, $E$ is the electric field, $ε$ is dielectric permittivity and $ε_0$ is dielectric permittivity of vacuum) has been accepted for pure NiO at ambient conditions, which has contributions from both electronic and ionic polarization [22]. Fuschillo, et al. [21] and Rao, et al. [22] studied the dielectric properties of the NiO ceramics in a wide range of frequencies as a function of temperature. Their investigation revealed a constant dielectric permittivity up to room temperature, which was expected as the polarization mechanisms in NiO are not sensitive to neither temperature nor frequency. It was shown that the Li doping considerably increases the dielectric permittivity of NiO [21]. Many reports confirmed the drastically increase of dielectric permittivity of NiO ceramics by introduction of the diverse doped atoms like titanium, lithium, cobalt, aluminum and sodium [18–20,23,24]. The investigations showed that point defects, secondary phases and grain boundaries could be responsible for this anomalous behavior.

Among the various approaches to induce different features into the materials, applying strain is one of the most interesting and suitable methods [1,2,4,5,8–10] . The attention to this new technique was attracted after the impressive developments in the thin-film technology in the beginning of 21$^{st}$ century. Many interesting phenomena emerged as a result of applying an appropriate strain to a material. A new research field of straintronics was formed [25]. However, up to this time, there is no systematic study of the effect of strain and intrinsic point defects on the dielectric properties of NiO thin films. Here, the growth of NiO films by pulsed laser deposition and their dielectric properties are investigated as a function of strain and defect concentration in a range of temperatures from 10 to 300 K.

## 2. Experimental
### 2.1. Substrate Preparation

SrTiO$_3$ (001) single-crystal substrates (Crystech) were used to grow nickel oxide thin films. Before starting the growth, the substrates were cleaned ultrasonically in acetone and methanol for 10 minutes in each, followed by an etching process for 30 second in BHF (pH~4.5) after a 5-minute soaking in DI water. An annealing process for 2.5 hours at 1000 ºC in air was done for all substrates to reach an atomically-flat step-terrace surface [26–29] . In case of Nb-doped STO substrates, the etching process was avoided, and after the washing substrates, the same annealing process was applied to them.

### 2.2. Film Deposition

Pulsed laser deposition (PLD) technique was used to prepare the nickel oxide films. The growth of single-phase crystalline NiO films by PLD method in a wide range of temperature from 10 to 750 ºC was already done recently [26]. For this purpose, an Nd-YAG pulsed laser (266 nm) operating at 10 Hz was used.

In this work, three different growth temperatures were selected, 100, 400 and 700 ºC. Before the deposition, the PLD chamber was pumped to a vacuum of 10$^{-8}$ Torr. To compensate for the oxygen deficiency in the deposited films, a high-purity oxygen gas was introduced into the chamber at the time of deposition, and the pressure was kept at the level of 15 mTorr. The substrate was set at the distance of 50 mm right above the NiO polycrystalline target used to deposit the material.

### 2.3. Characterization and dielectric measurements

To analyze the crystal structure of the films, an X-ray diffractometer was used operating at 40 kV 200 mA. Theta rocking curves around the detected Brag peaks were recorded. The film thickness was measured by X-ray reflectometer. To investigate surface morphology, atomic force microscopy in a non-contact mode has been used. The dielectric features of films, including capacitance and loss tangent, were measured using a precision LCR meter (Agilent E4980A) at different frequencies (20Hz to 2MHz) as a function of temperature from 10 to 300 K.

## 3. Results and Discussion

### 3.1. Structural characterization

Figure.1a shows the XRD spectra of the NiO films grown at different temperatures on $SrTiO_3$ (001) substrates. Appearance of a single (002)-NiO peak confirms the growth of pure nickel oxide and the absence of any other phases in the deposited material. Lowering the growth temperature leads to a shift of the NiO peak position to the left indicating the out-of-plane lattice expansion of the NiO structure, which might be due to the lattice mismatch between film and substrate. The theoretical lattice mismatch between NiO and $SrTiO_3$ is more than 5%, however, the relaxation of strain caused by this mismatch in the films starts at the initial stage of NiO growth, which was found in other works [26,30]. Two important parameters, which control strain accommodation are film thickness and growth temperature. Growing thicker film and increasing growth temperature lead to activation of the relaxation mechanisms and cause the suppression of the strain level in the film. The higher level of strain in the films grown at lower temperatures confirms important role of the thermal energy in the relaxation process. If growth temperatures is low, the lack of thermal driving force for the surface diffusion might partially decelerate the relaxation processes and a fraction of the mismatch strain may still remain in the film structure.

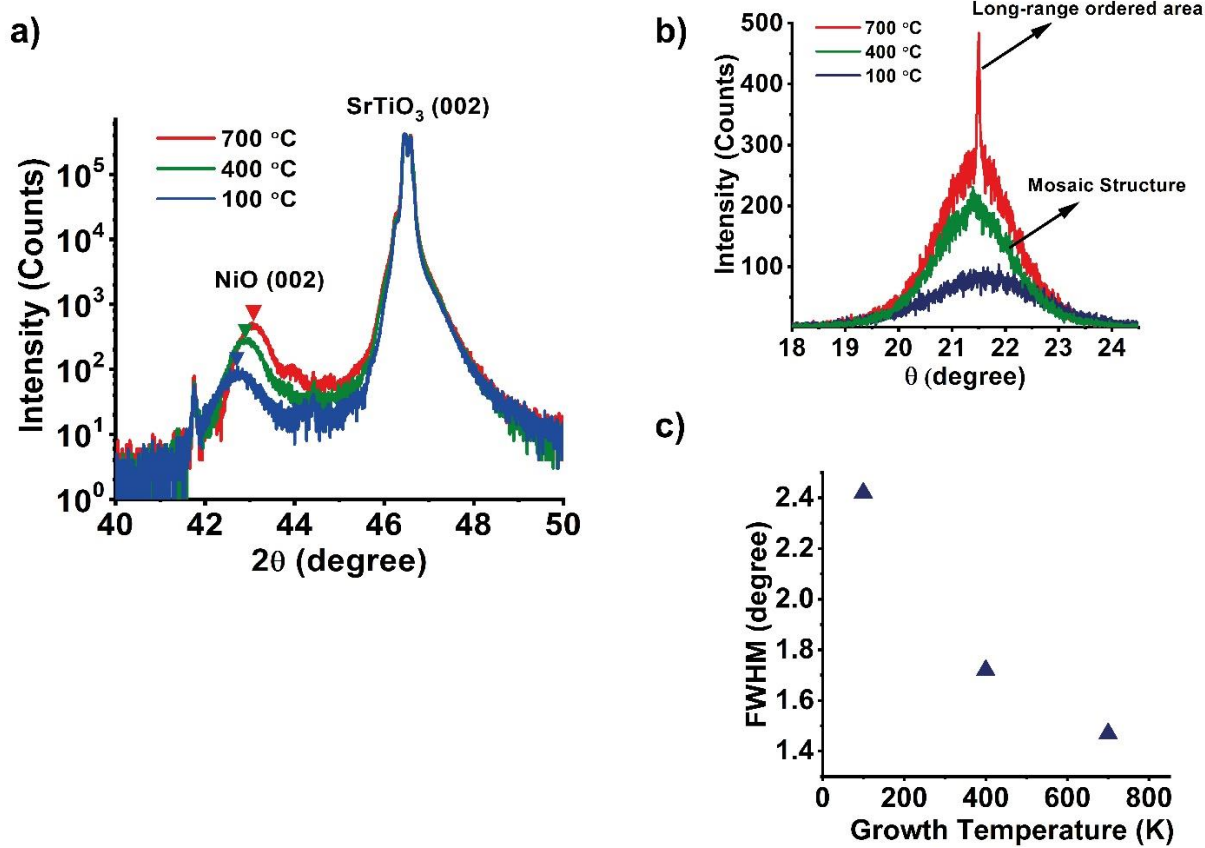

**Fig.1.** a) Two theta X-ray spectra of the NiO thin films grown on SrTiO3(001) substrates at different temperature, b) Theta rocking around (002)-NiO diffraction peak, c) Full width half maximum of the 002 NiO peak for the films grown at different temperatures.

The theta rocking curve around 002-NiO peak reveals an interesting crystalline feature. For the film grown at 750 °C it consists of two different components (see fig.1b). A sharp peak, which corresponds to an atomically long-range ordered area inside the film and a broad peak, which reflects a mosaic structure [26,31–35]. As the growth temperature decreases, the sharp peak disappears and the width of the broad peak increases (Fig. 1c). These two features imply that lowering the growth temperature results in larger number of defects in the film (increased mosaicity) that, in its turn, can significantly affect the electronic properties of the NiO. The clear fringes in XRR study (Fig. 2) for all three deposited films reveal the high quality interfaces (film surface and film/substrate interface), which cause the specular reflections of the incident beam.

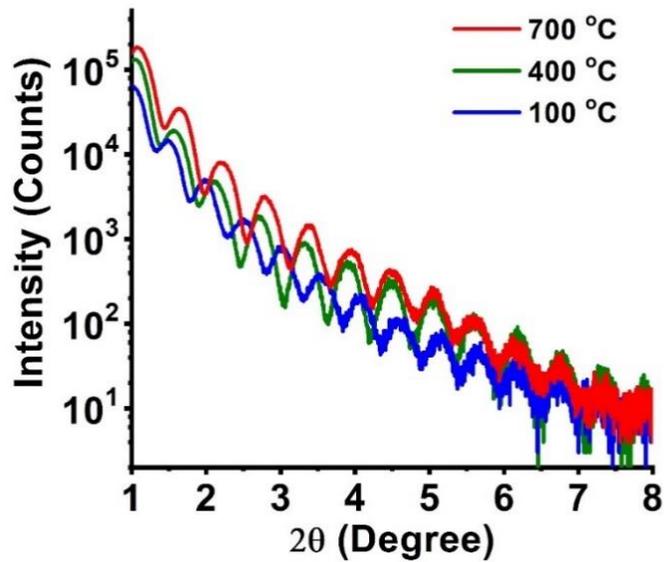

**Fig. 2.** X-ray reflectometry spectra of the NiO films grown at different temperatures on SrTiO3 (001) substrates.

### 3.2. Surface morphology

The study of surface by Atomic force microscopy confirms results of X-ray diffraction characterization. Figure.3 shows that lowering the growth temperature affects the surface morphology of the films. For the films deposited at 750 ºC, a surface with clear steps and terraces was detected and the step height is around 2 nm which is almost a half of unit-cell height of NiO. However, for the films grown at 400 and 100 ºC, the clear view of the steps gradually fades out. The degradation of the surface of the films grown at lower temperature might be related to the lack of thermal energy, which facilitates the surface diffusion and subsequently results in an atomically smooth surface.

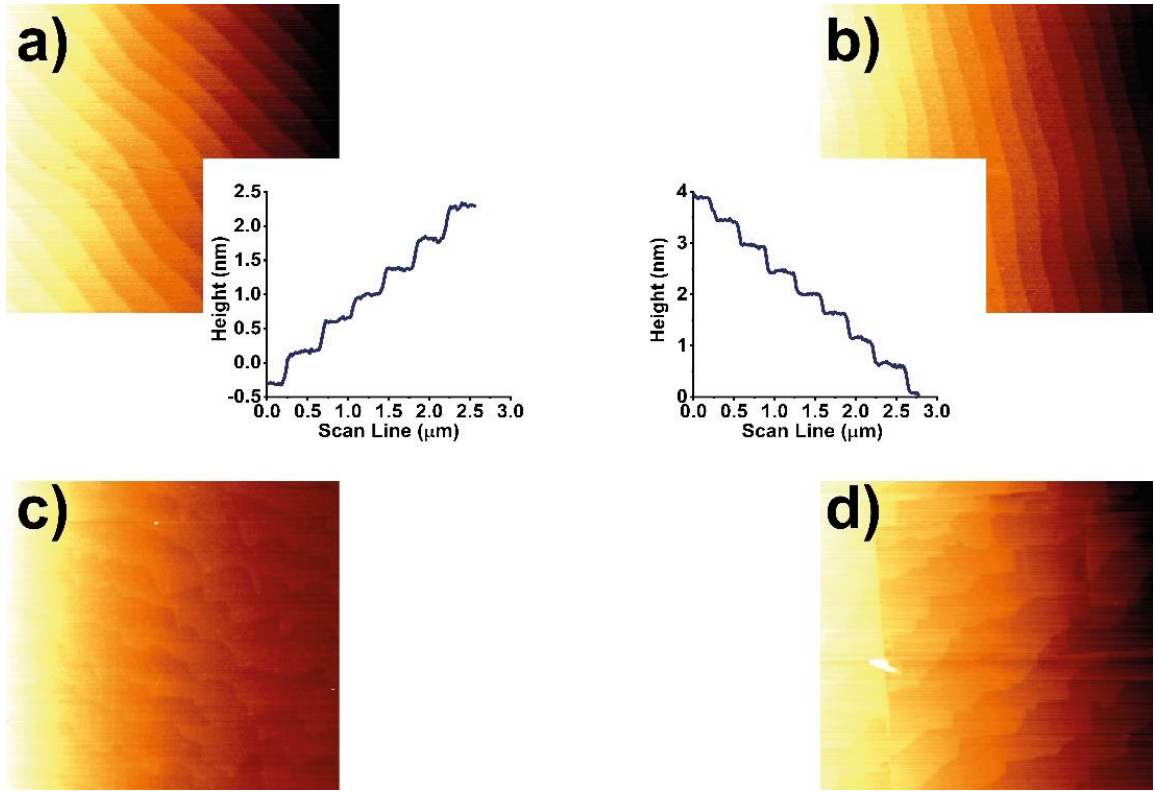

**Fig. 3.** Surface topographic images (5×5 µm$^2$) of a) SrTiO3 (001) surface after treatment; and NiO thin films grown at b) 700 °C, c) 400 °C, d) 100 °C.

### 3.3. Dielectric properties

To study the dielectric properties of NiO films, 0.5 wt% Nb-doped SrTiO$_3$ (001) conductive substrates were used as the bottom electrodes (see Fig. 4). An E-beam evaporation technique was employed to deposit Ti-Au electrode contacts on top of NiO film as shown in the figure. First, a 5-nm titanium film was deposited as an adhesive layer, then a 50-nm gold layer was deposited on the top of the contact. The contact structure was duplicated on bare part of the substrate to guarantee low contact resistance between electrodes and attached indium wires.

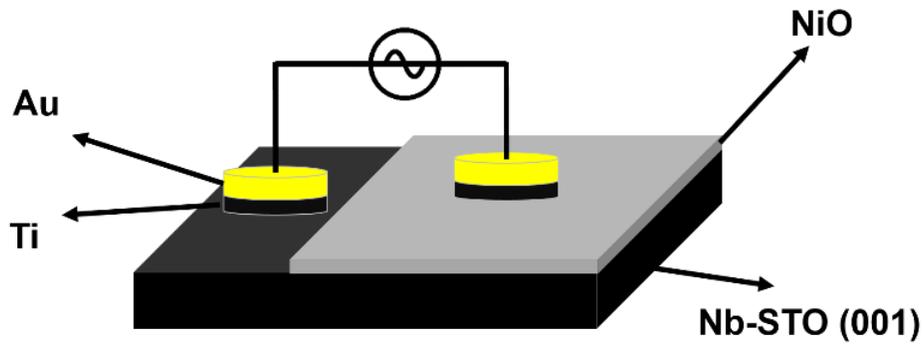

**Fig. 4.** The schematic image of Nb-STO/NiO/Ti/Au planar capacitor.

The capacitance of the films was measured in the temperature range from 10 to 300 K by applying a 1-mV AC voltage. Figure. 5 shows that the dielectric permittivity of NiO film grown at 100 ºC changes dramatically above 150 K, but the film grown at 750 ºC does not show any considerable change in the temperature range between 10 and 300 K as one expects for the NiO ionic structure. The film grown at 400 ºC shows a behavior between these limits.

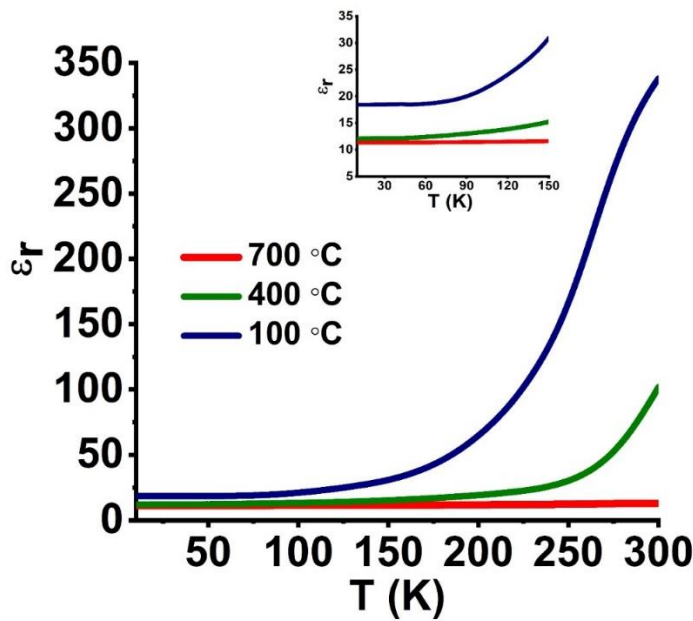

**Fig. 5.** The measured dielectric permittivity of the NiO films grown at different temperatures as a function of temperature at 1 kHz. The inset shows details of the dielectric behavior of the films at temperatures below 150 K.

These results, supported by the crystalline differences between the films, reveal the influence of defects on the dielectric properties of nickel oxide thin films. As was already discussed, lowering the growth temperature results in a defective structure (increased film mosaicity) which was concluded analyzing shape of the theta rocking curves. The formed defects offer a preferable place to accumulate charges and subsequently increase the high-temperature dielectric constant of the NiO film. It was already confirmed that the localized space charges in nickel oxide are thermally activated spices [36,37] which consequently increase the high temperature and low frequency dielectric constant of NiO. Increasing the growth temperature and achieving less defective film (atomically long-range-ordered film) dramatically decreases the high temperature dielectric constant, which is expected due to the decrease of the number of space charges in the film. Below 150 K (see the inset to fig. 5), the dielectric permittivity of all three samples does not show any abnormal behavior, and the curves become flat. Within this range of temperatures, the role of crystal lattice polarizability, which includes electronic and ionic components, becomes more dominant as compared to the space-charge polarizability. As a result, the effect of strain on the dielectric properties of NiO films becomes prominent. As the figure shows, the sample, which was grown at 100 ºC shows a dielectric permittivity, which is much higher than in those grown at 400 ºC and 750 ºC. This behavior indicates that the tensile strain in an ionic material can significantly increase the dielectric permittivity, as it was predicted in the first-principle calculations. This result directly comes from the properties of the lattice dynamics of the ionic binary oxides. The tensile strain decreases the frequency of the transverse optical phonon mode, which was predicted by Bousquet et al. [8] and BG Kim [9], which, subsequently, increases the low-frequency dielectric constant according to the Lyddane–Sachs–Teller relation [38]:

$$\frac{\omega_L^2}{\omega_T^2} = \frac{\epsilon(0)}{\epsilon(\infty)}, \qquad (1)$$

where $\omega_T$ and $\omega_L$ are the frequencies of the transverse and longitudinal optical phonon modes, respectively, $\varepsilon(0)$ is the low-frequency dielectric permittivity and $\varepsilon(\infty)$ is the high-frequency limit for electronic dielectric permittivity.

To investigate the nature of polarization in more detail, the dielectric properties of NiO grown at 100 °C were measured at lower frequencies. Figure 6 shows that at 20 Hz, the anomalous contribution of space-charges to the dielectric permittivity of the films starts from a lower temperature. This is predictable, as the activation of space-charge polarization is strongly dependent on the applied frequency compared to the electronic and ionic mechanisms. Below 100 K, both measured capacitance show same value and become independent of the frequency and temperature which indicates the dominant contribution of electronic and ionic polarizations in the dielectric permittivity of NiO film as these two mechanisms are not frequency and temperature dependent.

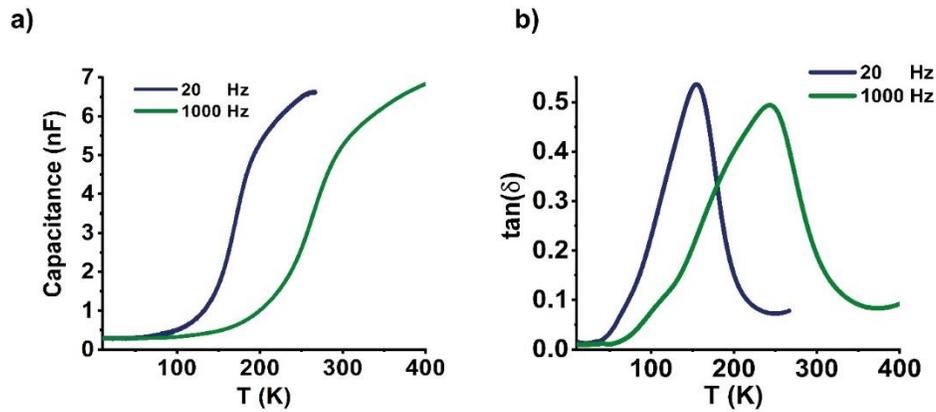

**Fig. 6.** The effect of the applied frequency on the temperature dependence of capacitance and loss factor of the NiO film grown at 100 °C.

## 4. Conclusion

Dielectric properties of the strained nickel oxide thin films have been studied as a function of temperature from 10 to 300 K. It was shown that the induced defects, and also strain play important role in the dielectric behavior of the NiO thin films. The contribution of space-charge polarization becomes dominant at higher temperatures, while the contribution of intrinsic electronic and ionic polarizations is more clear at temperatures below 100 K. The dramatic increase in the dielectric permittivity at high temperatures was attributed to the effect of space charges in the more defective films. At low temperatures, the strained films show some increase in the dielectric permittivity compared to the relaxed films, which can be interpreted based on the lattice dynamics properties of the cubic rocksalt binary oxides.

### Acknowledgement

This work was partially supported by National Research Foundation (NRF) of Korea (2015R1D1A1A02062239 and 2016R1A5A1008184) funded by the Korean Government.

# References

[1] D. G. Schlom, L.-Q. Chen, C.-B. Eom, K. M. Rabe, S. K. Streiffer, and J.-M. Triscone, Annu. Rev. Mater. Res. **37**, 589 (2007).

[2] J. Cao and J. Wu, Mater. Sci. Eng. R Reports **71**, 35 (2011).

[3] J. H. Haeni, P. Irvin, W. Chang, R. Uecker, P. Reiche, and Y. L. Li, **430**, 1 (2004).

[4] K. J. Choi, **1005**, 1005 (2010).

[5] D. G. Schlom, L. Q. Chen, C. J. Fennie, V. Gopalan, D. A. Muller, X. Pan, R. Ramesh, and R. Uecker, MRS Bull. **39**, 118 (2014).

[6] A. Biswas, M. Talha, A. Kashir, and Y. H. Jeong, Curr. Appl. Phys. **19,** 207 (2018).

[7] T. Archer, C. D. Pemmaraju, S. Sanvito, C. Franchini, J. He, A. Filippetti, P. Delugas, D. Puggioni, V. Fiorentini, R. Tiwari, and P. Majumdar, Phys. Rev. B - Condens. Matter Mater. Phys. **84**, 1 (2011).

[8] E. Bousquet, N. A. Spaldin, and P. Ghosez, Phys. Rev. Lett. **104**, 1 (2010).

[9] B. G. Kim, Solid State Commun. **151**, 674 (2011).

[10] X. Wan, H. C. Ding, S. Y. Savrasov, and C. G. Duan, Sci. Rep. **6**, 1 (2016).

[11] C. N. R. Rao and G. V. S. Rao, NBS Ref. Data Syst. Pap. (1974).

[12] P. Gupta, T. Dutta, S. Mal, and J. Narayan, J. Appl. Phys. **111**, 13706 (2012).

[13] E. Aytan, B. Debnath, F. Kargar, Y. Barlas, M. M. Lacerda, J. X. Li, R. K. Lake, J. Shi, and A. A. Balandin, Appl. Phys. Lett. **111**, 252402 (2017).

[14] A. G. Gavriliuk, I. A. Trojan, and V. V. Struzhkin, Phys. Rev. Lett. **109**, 1 (2012).

[15] X. B. Feng and N. M. Harrison, Phys. Rev. B - Condens. Matter Mater. Phys. **69**, 1 (2004).

[16] R. H. Kodama, S. A. Makhlouf, and A. E. Berkowitz, Phys. Rev. Lett. **79**, 1393 (1997).

[17] I. Sugiyama, N. Shibata, Z. Wang, S. Kobayashi, T. Yamamoto, and Y. Ikuhara, Nat. Nanotechnol. **8**, 266 (2013).

[18] Y. Lin, R. Zhao, J. Wang, J. Cai, C. W. Nan, Y. Wang, and L. Wei, J. Am. Ceram. Soc. **88**, 1808 (2005).

[19] S. Manna, K. Dutta, and S. K. De, J. Phys. D. Appl. Phys. **41**, (2008).

[20] Y. J. Hsiao, Y. S. Chang, T. H. Fang, Y. L. Chai, C. Y. Chung, and Y. H. Chang, J. Phys. D. Appl. Phys. **40**, 863 (2007).

[21] N. Fuschillo, B. Lalevic, and B. Leung, Thin Solid Films **24**, 181 (1974).

[22] K. V. Rao and A. Smakula, J. Appl. Phys. **36**, 2031 (1965).

[23] S. Manna and S. K. De, Solid State Commun. **150**, 399 (2010).

[24] J. Wu, C.-W. Nan, Y. Lin, and Y. Deng, Phys. Rev. Lett. **89**, 217601 (2002).

[25] A. A. Bukharaev, A. K. Zvezdin, A. P. Pyatakov, and Y. K. Fetisov, Uspekhi Fiz. Nauk **188**, 1288 (2018).

[26] A. Kashir, H.-W. Jeong, G. Lee, P. Mikheenko, and Y. H. Jeong, ArXiv E-Prints arXiv:1904.01780 (2019).

[27] M. Kawasaki, T. Maeda, R. Tsuchiya, and H. Koinuma, Science (80-. ). **266**, 1 (1993).

[28] A. Biswas, P. B. Rossen, C. H. Yang, W. Siemons, M. H. Jung, I. K. Yang, R. Ramesh, and Y. H. Jeong, Appl. Phys. Lett. **98**, 2009 (2011).

[29] A. Biswas, C. H. Yang, R. Ramesh, and Y. H. Jeong, Prog. Surf. Sci. **92**, 117 (2017).

[30] G. Chern and C. Cheng, J. Vac. Sci. Technol. A **17**, 1097 (1999).

[31] M. Becht, F. Wang, J. G. Wen, and T. Morishita, J. Cryst. Growth **170**, 799 (1997).

[32] O. Durand, A. Letoublon, D. J. Rogers, and F. Hosseini Teherani, Thin Solid Films **519**, 6369 (2011).

[33] P. F. Miceli, J. Weatherwax, T. Krentsel, and C. J. Palmstrøm, Phys. B Condens. Matter **221**, 230 (1996).

[34] A. R. Wildes, J. Mayer, and K. Theis-Bröhl, Thin Solid Films **401**, 7 (2001).

[35] H. Search, C. Journals, A. Contact, M. Iopscience, T. Table, and I. P. Address, **2669**, (1999).

[36] K. Oka, T. Yanagida, K. Nagashima, H. Tanaka, and T. Kawai, J. Appl. Phys. **104**, (2008).

[37] V. Usha, S. Kalyanaraman, R. Vettumperumal, and R. Thangavel, Phys. B Condens. Matter **504**, 63 (2017).

[38] P. Andrade and S. Porto, Brazillian J. Phys. **3**, 337 (1973).